\documentclass[12pt]{article}
\pdfoutput=1
\usepackage{jheppub}
\usepackage{mathtools}

\def\be{\begin{equation}}
\def\ee{\end{equation}}
\def\ba#1\ea{\begin{align}#1\end{align}}
\def\bg#1\eg{\begin{gather}#1\end{gather}}
\def\bm#1\em{\begin{multline}#1\end{multline}}
\def\bmd#1\emd{\begin{multlined}#1\end{multlined}}

\def\a{\alpha}
\def\b{\beta}

\def\d{\delta}

\def\e{\epsilon}
\def\ve{\varepsilon}
\def\g{\gamma}

\def\h{\eta}

\def\l{\lambda}

\def\m{\mu}
\def\n{\nu}
\def\p{\phi}

\def\r{\rho}
\def\s{\sigma}
\def\t{\tau}

\def\la{\label}
\def\ci{\cite}
\def\re{\ref}
\def\er{\eqref}
\def\se{\section}
\def\sse{\subsection}
\def\ssse{\subsubsection}
\def\fr{\frac}
\def\na{\nabla}
\def\pa{\partial}
\def\td{\tilde}

\def\ph{\phantom}
\def\eq{\equiv}
\def\nn{\nonumber}
\def\qu{\quad}

\def\lt{\left}
\def\rt{\right}
\def\({\left(}
\def\){\right)}
\def\[{\left[}
\def\]{\right]}
\def\lra{\leftrightarrow}

\def\Tr{{\rm Tr}}
\def\mo{{\mathcal O}}
\def\zb{{\bar z}}
\def\tdd{{\td\d}}
\def\tdr{{\td\r}}
\def\tdt{{\td\t}}

\def\Area{{\rm Area}}
\def\inside{{\rm inside}}
\def\outside{{\rm outside}}
\def\total{{\rm total}}
\def\sym{{\rm sym}}

\newcommand{\T}[3]{{#1^{#2}_{\ph{#2}#3}}}
\newcommand{\Tu}[3]{{#1_{#2}^{\ph{#2}#3}}}

\begin{document}

\subheader{SU-ITP-13/21}
\title{Holographic Entanglement Entropy for \\General Higher Derivative Gravity}
\author{Xi Dong}
\affiliation{Stanford Institute for Theoretical Physics, Department of Physics, Stanford University, Stanford, CA 94305, U.S.A.}
\emailAdd{xidong@stanford.edu}

\abstract{We propose a general formula for calculating the entanglement entropy in theories dual to higher derivative gravity where the Lagrangian is a contraction of Riemann tensors.  Our formula consists of Wald's formula for the black hole entropy, as well as corrections involving the extrinsic curvature.  We derive these corrections by noting that they arise from naively higher order contributions to the action which are enhanced due to would-be logarithmic divergences.  Our formula reproduces the Jacobson-Myers entropy in the context of Lovelock gravity, and agrees with existing results for general four-derivative gravity.

We emphasize that the formula should be evaluated on a particular bulk surface whose location can in principle be determined by solving the equations of motion with conical boundary conditions.  This may be difficult in practice, and an alternative method is desirable.  A natural prescription is simply minimizing our formula, analogous to the Ryu-Takayanagi prescription for Einstein gravity.  We show that this is correct in several examples including Lovelock and general four-derivative gravity.}
\maketitle

\se{Introduction}
The study of black hole thermodynamics has lead to a remarkably simple formula for the gravitational entropy associated with horizons \cite{Bekenstein:1973ur, Bardeen:1973gs, Hawking:1974sw}:
\be
S = \fr{\Area}{4G_N} \,.
\ee
Gibbons and Hawking found a method of calculating the gravitational entropy by studying Euclidean gravity solutions with a $U(1)$ symmetry \cite{Gibbons:1976ue}.  In the context of gauge-gravity duality, this simple formula was elegantly generalized by Ryu and Takayanagi to the entanglement entropy in field theories with holographic duals \cite{Ryu:2006bv}.  In this case the area is evaluated on the minimal surface in the dual bulk geometry that is anchored to the boundary of the spatial region of interest.  For spherical regions in holographic CFTs, there is a $U(1)$ symmetry allowing us to map the entanglement entropy to the horizon entropy of hyperbolic black holes \cite{Casini:2011kv}, but in the general case no $U(1)$ symmetry exists.

Recently, Lewkowycz and Maldacena generalized the Euclidean method of calculating the gravitational entropy to cases without a $U(1)$ symmetry \cite{Lewkowycz:2013nqa}.  Using this and the replica trick, they proved the Ryu-Takayanagi conjecture\footnote{There is an assumption about the replica symmetry in the bulk, which we will come back to later.}.

It is natural to ask for a generalization of the Ryu-Takayanagi prescription to general theories of higher derivative gravity\footnote{After all, string theory predicts such $\a'$-corrections.  Note that another natural question involves higher-loop corrections which are not subjects of this paper but were analyzed in \ci{Barrella:2013wja, Faulkner:2013ana}.}.  In the black hole context the analogous question was answered by Wald who proposed the following entropy formula \cite{Wald:1993nt, Iyer:1995kg, Jacobson:1993vj}:
\be\la{wald}
S_{\rm Wald} = -2\pi \int d^d y \sqrt g \fr{\pa L}{\pa R_{\m\r\n\s}} \ve_{\m\r} \ve_{\n\s} \,,
\ee
where notations are explained in Appendix~\re{appnt}.  One might guess that Wald's formula also serves as the prescription for the entanglement entropy in theories dual to higher derivative gravity.  However, this cannot be correct because it would give wrong universal terms in the entanglement entropy \cite{Hung:2011xb}.

On the other hand, there exists a different formula for the black hole entropy in Lovelock gravity \cite{Lovelock:1971yv, raey}, which was derived using a Hamiltonian approach by Jacobson and Myers \cite{Jacobson:1993xs}.  It differs from Wald's formula only by terms involving the extrinsic curvature, which vanishes at a Killing horizon.  However, their differences matter if we use them for the entanglement entropy, as minimal surfaces (or their analogs) generally have nonzero extrinsic curvature.  Interestingly, the Jacobson-Myers formula used as the holographic entanglement entropy gives the correct universal terms for Gauss-Bonnet gravity \cite{Hung:2011xb}.

This leaves us with two natural questions.  First, does the Jacobson-Myers formula work in general Lovelock gravity?  Second, is there an entropy formula which works for general higher derivative gravity?

In this paper, we propose the following formula for calculating the holographic entanglement entropy in a general theory dual to higher derivative gravity where the Lagrangian $L(R_{\m\r\n\s})$ is built from arbitrary contractions of Riemann tensors:
\be\la{eei}
S_{EE}= 2\pi\int d^d y \sqrt{g} \lt\{ \fr{\pa L}{\pa R_{z\zb z\zb}} + \sum_\a \(\fr{\pa^2 L}{\pa R_{zizj} \pa R_{\zb k\zb l}}\)_\a \fr{8K_{zij} K_{\zb kl}}{q_\a+1} \rt\} \,.
\ee
For a full explanation of notations we refer our readers to Section~\re{secee} and Appendix~\re{appnt}.  We briefly mention how to calculate the second term here.  In the second derivative of $L$ we expand the following components of the Riemann tensor in terms of the extrinsic curvature $K_{aij}$, $Q_{abij} \eq \pa_a K_{bij}$, and the lower-dimensional Riemann tensor $r_{ikjl}$:
\ba
R_{abij} &= \td R_{abij} + g^{kl} (K_{ajk} K_{bil} - K_{aik} K_{bjl}) \,,\nn\\
R_{aibj} &= \td R_{aibj} + g^{kl} K_{ajk} K_{bil} - Q_{abij} \,,\la{rexpi}\\
R_{ikjl} &= r_{ikjl} + \hat g^{ab} (K_{ail} K_{bjk} - K_{aij} K_{bkl}) \,.\nn
\ea
The definitions of $\td R_{abij}$ and $\td R_{aibj}$ are in \er{rtd} but not required here.  Let us use $\a$ to label the terms in the expansion.  For each term (which is a product) we define $q_\a$ as the number of $Q_{zzij}$ and $Q_{\zb\zb ij}$, plus one half of the number of $K_{aij}$, $R_{abci}$, and $R_{aijk}$.  Finally we sum over $\a$ with weights $1/(1+q_\a)$.  We can then eliminate $\td R_{abij}$, $\td R_{aibj}$, and $r_{ikjl}$ (if we want) using \er{rexpi}, arriving at an expression involving only components of $R_{\m\n\r\s}$, $K_{aij}$ and $Q_{abij}$.  The expansion and resummation can be thought of as a simple prescription to generate higher order terms in $K_{aij}$ and $Q_{abij}$.

An equivalent but covariant form of the formula is
\bm\la{eeci}
\!\!\!\!\!
S_{EE}= 2\pi\int d^d y \sqrt{g} \lt\{ -\fr{\pa L}{\pa R_{\m\r\n\s}} \ve_{\m\r} \ve_{\n\s} + \sum_\a \(\fr{\pa^2 L}{\pa R_{\m_1\r_1\n_1\s_1} \pa R_{\m_2\r_2\n_2\s_2}}\)_\a \fr{2K_{\l_1\r_1\s_1} K_{\l_2\r_2\s_2}}{q_\a+1} \times \rt. \\
\lt.\phantom{\frac12} \times \[ (n_{\m_1\m_2} n_{\n_1\n_2}-\ve_{\m_1\m_2} \ve_{\n_1\n_2}) n^{\l_1\l_2} + (n_{\m_1\m_2} \ve_{\n_1\n_2}+\ve_{\m_1\m_2} n_{\n_1\n_2}) \ve^{\l_1\l_2}\] \rt\} \,.
\em
Here $n_{\m\n}$ and $\ve_{\m\n}$ reduces to the metric and Levi-Civita tensor in the two orthogonal directions with all other components vanishing.

Our entropy formula consists of Wald's formula and corrections involving the extrinsic curvature.  We derive this formula by a generalization of the Euclidean method involving regularized squashed cones.  From the derivation we see that the extrinsic curvature terms arise from a subtlety not present in Einstein gravity, which is that a naively higher order contribution to the action may be enhanced due to a would-be logarithmic divergence.  These extrinsic curvature terms can therefore be thought of as ``anomalies'' in the variation of the action.  From this perspective, the number $q_\a$ is analogous to an ``anomaly coefficient'' that we associate to each term in the expansion of $\fr{\pa^2 L}{\pa R_{zizj} \pa R_{\zb k\zb l}}$ mentioned above.  We postpone the details until Section~\re{secee}.

We find that our formula \er{eei} fully reproduces the Jacobson-Myers entropy for general Lovelock gravity.  This is a nontrivial check for our formula because it requires all projected and extrinsic curvature terms sum into intrinsic curvature terms with the correct coefficients.  Another special case is in the context of general four-derivative gravity, where our formula reproduces a recent result in \cite{Fursaev:2013fta}.

We emphasize that our formula \er{eei} should be evaluated on a particular codimension 2 surface in the bulk whose location can in principle be determined by solving all bulk equations of motion with conical boundary conditions.  This is well-defined but may be difficult in practice, and an alternative method of finding the location of the surface is desirable.  A natural conjecture is that it is determined by minimizing the same entropy formula \er{eei}, analogous to the Ryu-Takayanagi prescription in Einstein gravity.  We show that this is true in three examples including Lovelock and general four-derivative gravity.  We leave a general solution to this problem for future work.

Even though we focus on the holographic entanglement entropy in this paper, our formula \er{eei} can also be used for the black hole entropy\footnote{In short this is because our derivation of the formula builds upon the generalized gravitational entropy method of \cite{Lewkowycz:2013nqa}.}, in which case it is evaluated on the horizon.  In the special case where it is a Killing horizon, our argument can be thought of as a holographic derivation of Wald's formula.

We begin in Section~\re{seclm} with a review on the derivation of the Ryu-Takayanagi prescription \cite{Lewkowycz:2013nqa}, rephrasing it sometimes for our later generalization beyond Einstein gravity.  In Section~\re{secfm}, we derive our entropy formula \er{eei}.  We first point out a subtlety not present in Einstein gravity due to would-be logarithmic divergences, and then introduce the regularized squashed cone method in Section~\re{secrsc}.  After deriving the general formula in Section~\re{secee}, we apply it to several examples in Section~\re{secex} and confirm that it agrees with existing results.  In Section~\re{secmin}, we investigate another part of the story, i.e.\ whether minimizing our formula gives the correct surface on which we should evaluate it.  We first confirm this in two ways in Lovelock gravity, and then work towards a general derivation.  Our general derivation is not yet complete, but it suffices for several examples including general four-derivative gravity.  We conclude with open questions and future directions in Section~\re{seccon}.  In Appendix~\re{appnt} we summarize our notations and conventions for quick reference, and in Appendix~\re{appsc} we derive a few details about squashed cones.

\se{Review: derivation of the Ryu-Takayanagi prescription}\la{seclm}

In field theories dual to Einstein gravity, the entanglement entropy of a spatial region is given by the area of the minimal surface in the bulk:
\be\la{rt}
S_{EE} = \frac{\Area_{\min}}{4G_N} \,.
\ee
Here $G_N$ is the bulk Newton's constant, and the surface is found by minimizing its area among all bulk surfaces homologous to the spatial region under consideration in the field theory.  This prescription was conjectured by Ryu and Takayanagi \ci{Ryu:2006bv} and proven by Lewkowycz and Maldacena \ci{Lewkowycz:2013nqa}.  Proof of the formula for special cases includes \ci{Casini:2011kv, Hartman:2013mia, Faulkner:2013yia}.  Earlier attempts to prove the formula include \cite{Fursaev:2006ih, Headrick:2010zt}.  One-loop corrections were analyzed in \ci{Barrella:2013wja, Faulkner:2013ana}.  A covariant version of the prescription which applies in general time-dependent cases was proposed in \cite{Hubeny:2007xt}.

Let us review the derivation of the Ryu-Takayanagi prescription.  The basic idea is to use the replica trick and extend it to the bulk.  Recall that the R\'enyi entropy is defined by
\be
S_n= -\fr{1}{n-1} \log \Tr[\r^n] \,,
\ee
where $\r$ is the reduced density matrix associated with the subsystem $A$ under consideration.  We obtain the entanglement entropy $S_{EE}=-\Tr[\r\log\r]$ by analytically continuing the R\'enyi entropy to $n\to1$.  The R\'enyi entropy may be computed from
\be\la{renyi}
S_n= -\frac{1}{n-1}(\log Z_n-n\log Z_1) \,,
\ee
where $Z_n$ (with integer $n>1$) is the partition function of the field theory on a suitable manifold $M_n$ known as the $n$-fold cover.  In particular, $M_1$ is the original spacetime manifold (analytically continued to Euclidean signature), and $M_n$ is defined by taking $n$ copies of $M_1$, cutting each of them apart at the spatial region $A$, and gluing them together in a cyclic order.  In terms of the $\t$ coordinate defined locally as the angle around the boundary $\pa A$ of $A$,  this procedure can be thought of as extending the range of $\t$ from $2\pi$ to $2\pi n$.  An example of the $n$-fold cover is shown on the left of Figure~\re{figorb}.

\begin{figure}[h]
\centering
\raisebox{-0.5\height}{\includegraphics[width=0.3\textwidth]{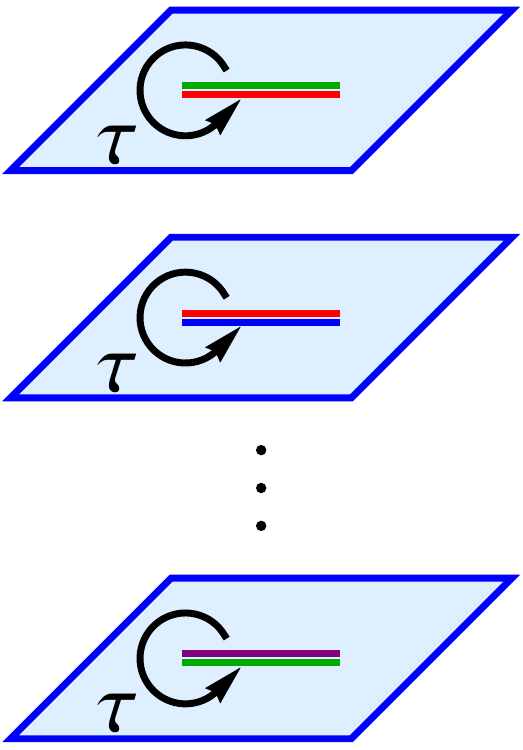}}
\qu\raisebox{-0.5\height}{\Huge$\xRightarrow{Z_n}$}\qu
\raisebox{-0.5\height}{\includegraphics[width=0.3\textwidth]{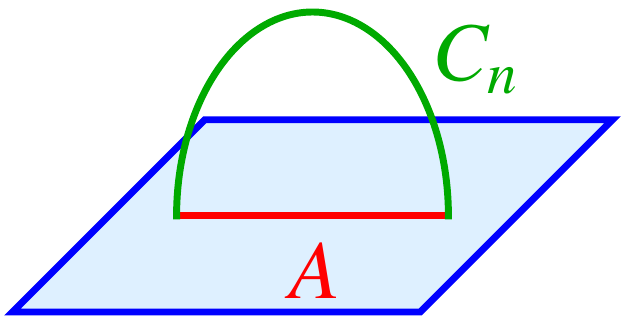}}
\caption{\label{figorb}Left: the $n$-fold cover $M_n$ for a $1+1$ dimensional field theory.  Right: the $Z_n$ orbifold which has a conical defect $C_n$ in the bulk dual.}
\end{figure}

For theories with a holographic dual one can build a suitable bulk solution\footnote{In general there may be more than one bulk solutions (or saddle points).  Here let us choose the dominant saddle.} $B_n$ whose boundary is $M_n$.  The gauge-gravity duality \ci{Maldacena:1997re, Gubser:1998bc, Witten:1998qj} identifies the field theory partition function on $M_n$ with the on-shell bulk action on $B_n$ in the large-$N$ limit:
\be\la{adscft}
Z_n\eq Z[M_n]= e^{-S[B_n]}+\cdots \,,
\ee                                                       
where the dots denote corrections from both $1/N$ effects and subdominant saddles.

When $n$ is not an integer, $Z_n$ in general can no longer be written as a partition function with a local action\footnote{A notable exception is when there is a $U(1)$ rotational symmetry in $\t$, in which case it makes sense to extend its range to any real value.}.  In particular, we cannot in general hope to analytically continue $M_n$ to non-integer $n$.  We could still calculate $Z[M_n]$ for integer $n$ first and then analytically continue the result, but this might be technically difficult (but possible in certain cases such as AdS$_3$/CFT$_2$ \ci{Hartman:2013mia, Faulkner:2013yia}).

The key observation in \ci{Lewkowycz:2013nqa} is that the dual side provides a much ``better'' analytic continuation -- we may analytically continue (an orbifold of) $B_n$ to suitable bulk configurations at non-integer $n$.  To review the argument, let us first note that the boundary $M_n$ at integer $n$ has a $Z_n$ symmetry that cyclically permutes the $n$ replicas.  Let us assume that this replica symmetry extends to the bulk $B_n$ for the dominant saddle.  We may then consider the orbifold
\be
\hat B_n = B_n / Z_n \,,
\ee
which is regular except at the fixed points of the $Z_n$ action.  The fixed points form a codimension 2 surface\footnote{This codimension 2 surface may be disconnected as is the case for certain choices of the spatial region $A$.} with a conical defect in the bulk and end on $\pa A$ at the boundary.  One way to see this is the following.  Note that the boundary of $\hat B_n$ is precisely $M_n/Z_n = M_1$, the original manifold.  The $Z_n$ symmetry acts on the boundary $B_n$ as $\t\to\t+2\pi$, where $\t$ as defined before is the angle around $\pa A$.  Therefore the fixed points on the boundary are precisely $\pa A$.  We can extend the $\t$ coordinate locally into the bulk such that the $Z_n$ replica symmetry acts in the same way.  Therefore the whole set of fixed points must form a codimension 2 surface ending on $\pa A$.  Let us call this codimension 2 surface $C_n$.  Since $B_n$ is a bulk solution (for integer $n$) and must therefore be regular in the interior, its $Z_n$ orbifold has a conical defect at $C_n$ with opening angle $2\pi/n$ (or deficit $2\pi-2\pi/n$).  An example of the orbifold is shown on the right of Figure~\re{figorb}.

How does this help us calculate the entanglement entropy?  We note that by construction, we may write
\be\label{sorb}
S[B_n] = n S[\hat B_n]
\ee
at integer $n$, where we stress that $S[\hat B_n]$ is the classical action for the bulk configuration $\hat B_n$ not including any contribution from the conical defect.  In particular, there is no Gibbons--Hawking--York (GHY) surface term at $C_n$.  This is because we would like \er{sorb} to hold, and $S[B_n]$ certainly does not contain any significant contribution at $C_n$.  From this it is also easy to see that at the asymptotic boundary $M_1$, we do need to include in $S[\hat B_n]$ the usual GHY surface term as well as counterterms.

If we can somehow analytically continue $\hat B_n$ to non-integer $n$ (which we will do in the next subsection), then we could use \er{sorb} to define $S[B_n]$.  Plugging it to \er{renyi} and \er{adscft}, we find
\be\label{snbh}
S_n= \frac{n}{n-1}\(S[\hat B_n]-S[\hat B_1]\) \,,
\ee
where $\hat B_1=B_1$ is simply the original bulk dual.  We may then expand around $n=1$ and get the entanglement entropy.

\sse{Two methods}\la{sectm}

There are two equivalent ways of finding the analytic continuation of $\hat B_n$.  We will call them the ``boundary condition'' method and the ``cosmic brane'' method.  In both methods, the analytic continuation of $\hat B_n$ still has a conical defect at some codimension 2 surface $C_n$ with opening angle $2\pi/n$, even though they can no longer be thought of as an $Z_n$ orbifold of some regular geometry.  As we will see $C_n$ approaches the minimal surface in the Ryu-Takayanagi proposal as $n$ goes to 1.  Adopting a suitable set of local coordinates in which $\r$ parameterizes the minimal distance to $C_n$, the metric may be written as
\be\label{met1}
ds^2 = \r^{-2\e} (d\r^2 + \r^2 d\t^2) + (g_{ij} + 2K_{aij} x^a) dy^i dy^j + \cdots \,,
\ee
where we use $a$, $b$, $\cdots$ as the indices in the $(\r,\t)$ plane orthogonal to $C_n$, and use $i$, $j$, $\cdots$ as the indices along $C_n$.  The angular coordinate $\t$ has a range of $2\pi$, and it is easy to see that the conical deficit of the above metric at $\r=0$ is $2\pi\e$.  For this to agree with the required deficit at integer $n$, we set
\be
\e = 1-\fr1n \,.
\ee
$K_{aij}$ is the extrinsic curvature tensor\footnote{To avoid possible confusion we point out that the extrinsic curvature $K_{aij}$ is often defined with the index $a$ labeling an orthonormal basis.  We do not require that here.} of the codimension 2 surface $C_n$, which is sometimes also written as $K_{(a)ij}$ or $_{(a)}K_{ij}$ in the literature.  In our coordinates it may be defined as $K_{aij} \eq \fr12 \pa_a G_{ij}$.  The dots in \er{met1} denotes higher powers of $\r$ which are subleading near $C_n$.  The form of these corrections as well as the justification for \er{met1} are worked out in details in Appendix~\re{appsc}.  This geometry is also known as the squashed cone due to its lack of $U(1)$ symmetry \ci{Fursaev:2013fta}.

Let us first review the boundary condition method \ci{Lewkowycz:2013nqa}.  Here we find $\hat B_n$ by simply solving all bulk equations of motion with the boundary condition that the metric should behave like $\er{met1}$ near some codimension 2 surface $C_n$ ending on $\pa A$.  This can be thought of as an unconventional ``IR'' boundary condition.  We impose the usual UV boundary condition in gauge-gravity duality, noting that the UV boundary of $\hat B_n$ is always $M_1$.  If we have additional matter fields $\p$, their boundary condition near $\r=0$ is $\p=\p_0+\p_a x^a +\cdots$ with $\p_0$, $\p_a$ generically finite as $\r\to0$ (in the complex basis defined below).  We derive these boundary conditions for the metric and matter fields and make them more precise in Appendix~\re{appsc}.  In general we need to impose as many boundary conditions as required by the equation of motion.

It turns out that this prescription can be used in a simple way to fix the location of $C_n$ in the $n\to1$ limit.  Going to complex coordinates $z=\r e^{i\t}$, the $zz$ component of the Einstein equation has a term that potentially diverges as $\r\to0$:
\be\la{rzz}
R_{zz} = 2K_z \fr{\e}{z} + \cdots \,,
\ee
where $K_a \eq K_{aij}g^{ij}$ is the trace of the extrinsic curvature, and the dots denote terms that are less divergent as $\r\to0$ or higher order in $\e$.  The stress tensor from the matter sector is not expected to diverge -- it is certainly regular at integer $n$ because the unorbifolded solution $B_n$ is regular.  Therefore we conclude that the $1/z$ divergence in \er{rzz} must vanish:
\be\la{kz}
K_z = 0
\ee
in the $n\to1$ limit.  This is precisely the equation for a minimal\footnote{We ignore extremal but non-minimal surfaces here because they correspond to subdominant saddles.} surface.

Equivalently, one can use the cosmic brane method.  We note that the conical defect at $C_n$ may be reproduced by an a codimension 2 brane with a suitable tension.  We are then lead to the prescription that $\hat B_n$ is found by solving all equations of motion resulting from the combined action\footnote{In particular, the equations of motion for the embedding coordinates guarantee that the brane is ``straight'' \ci{Unruh:1989hy, Boisseau:1996bp}.  This ensures that when $n$ is an integer, we may take $n$ copies of $\hat B_n$ and smoothly glue them together along the location of the brane, thus reproducing the parent space $B_n$.}
\be
S_{\rm total} = S_{EH} + S_B = -\fr{1}{16\pi G_N} \int d^Dx \sqrt G R + \fr{\e}{4G_N} \int d^dy \sqrt g \,,
\ee
where $D$ is the total number of bulk dimensions, and $d=D-2$ is the dimension of $C_n$ or the brane.  We use $G_{\mu\nu}$ and $R$ to denote the metric and Ricci scalar in $D$ dimensions, and use $g_{ij}$ to denote the induced metric in $d$ dimensions.  We will consistently use capital letters for higher dimensional quantities and lower-case letters for intrinsic (as opposed to projected) quantities in lower dimensions.

In the cosmic brane picture, it is clear that in the $n\to1$ or $\e\to0$ limit, we can treat it as a probe brane and find its location by minimizing its action $S_B$ without it backreacting on the bulk fields.  This gives precisely the minimal surface.  This is equivalent to the boundary condition method since the cosmic brane imposes the correct boundary condition at $C_n$.  We stress that the cosmic brane is an auxiliary tool for finding the correct analytically continued $\hat B_n$, and once we find it we should not include its tension in $S[\hat B_n]$.

Therefore, both methods give the minimal surface as the location of $C_n$ as $n\to1$.  Before completing the derivation of the Ryu-Takayanagi prescription, we note that both methods in principle determine $\hat B_n$ for any real $n\ge1$, not just near $n=1$.  In particular, they reproduce the correct $\hat B_n=B_n/Z_n$ from the replica trick when $n$ is an integer.

\sse{Variation of the action}\la{secva}
We need one more step to finish the proof of the Ryu-Takayanagi prescription.  We need to calculate the variation of the action $S[\hat B_n]$ to linear order in $n-1$ to get the entanglement entropy, because taking the $n\to1$ limit of \er{snbh} we get
\be
S_{EE} = \lim_{n\to1} \frac{n}{n-1}\(S[\hat B_n]-S[\hat B_1]\)  = \lt. \pa_n S[\hat B_n]\rt|_{n=1} \,.
\ee
For Einstein gravity this can be argued to be the area of $C_1$ divided by $4G_N$ in the following way.  We are varying away from $n=1$, a solution to the equations of motion with no conical defect, and therefore the only contributions to the variation of the action are boundary terms at $C_1$.  We have boundary terms because we do not include any contribution from $C_1$, and this means that we should excise a small region around $C_1$, introducing a boundary.  An explicit calculation of the boundary terms from the Einstein-Hilbert action gives \cite{Lewkowycz:2013nqa}
\be
S_{EE} = \fr{\Area(C_1)}{4G_N} \,.
\ee
We will not repeat the calculation here, because for higher derivative gravity we will use the regularized squashed cone method to be reviewed in Section~\re{secrsc}.  Combined with the fact that $C_1$ coincides with the minimal surface, this prove the Ryu-Takayanagi prescription \er{rt}.


\se{A general entropy formula}\la{secfm}

In this Section~we derive an entropy formula for holographic theories dual to general higher derivative gravity.

Before going into the details, let us first point out a new feature.  In the previous case of Einstein gravity, as we mentioned in Section~\re{secva} the $\mo(n-1)$ variation of the action is purely a boundary term due to the equations of motion.  This is not necessarily true for higher derivative gravity because the integral over $\r$ could potentially be divergent at $n=1$, enhancing certain contributions that would naively be higher order in $n-1$ to the linear order.

Let us see this explicitly in the simplest nontrivial example.  Consider the four-derivative theory
\be
L = R_{\m\r\n\s} R^{\m\r\n\s} \,.
\ee
The following term from the Lagrangian
\be
R_{zizj} R^{zizj} = 4 \fr{\e^2}{z\zb} K_{zij} \Tu K \zb{ij} (G^{z\zb})^2 + \cdots
\ee
appears to be of order $\e^2$ and one might have thought that it should not contribute to the linear order.  However, this term scales like $1/\r^2$ in the $\e\to0$ limit, and the integral over $\r$ has a potential logarithmic divergence.  In such cases, we have to keep all powers of $\r^\e$ in the integrand, because the would-be logarithmic divergence becomes a $1/\e$ enhancement:
\be
\int \r d\r \fr{1}{\r^2} \r^{\b\e} \sim \fr{1}{\b\e} \,.
\ee
In this particular example, $\b=2$ as can be determined from $\sqrt G$ and $(G^{z\zb})^2$.  Note that at large $\r$ (i.e.\ far away from the conical defect) there are corrections which regulate the behavior of the integrand in the above equation.

This is analogous to how $1/\e$ poles arise in dimensional regularization.  There we analytically continue in $\e=4-D$ (for a 4-dimensional field theory).  Here we have a very similar situation with $\e=1-1/n$.

Due to this new feature, we have to keep certain terms that are of order $\e^2$ in the Lagrangian.  In practice the calculation is much simpler and cleaner if we use the regularized squashed cone method.  We will review it next, and then use it to derive the general entropy formula.

\sse{Regularized squashed cones}\la{secrsc}

Let us review the regularized squashed cone method.  We call a cone symmetric if it is invariant under $U(1)$ rotations, and squashed if it is not.  In the symmetric case the Euclidean method, also known as the cone method, has been discussed in various forms \cite{Bardeen:1973gs, Banados:1993qp, Carlip:1993sa, Susskind:1994sm, Fursaev:1995ef, Nelson:1994na}, whereas cases without a $U(1)$ symmetry was discussed in \cite{Lewkowycz:2013nqa, Fursaev:2013fta}.  In the rest of the paper we will often refer to squashed cones simply as cones, as we focus on cases without a $U(1)$ symmetry.

We would like to calculate the variation of the action $S[\hat B_n]$ to linear order in $n-1$ or $\e$:
\be\la{seeo}
S_{EE} = \lt. \pa_\e S[\hat B_n]\rt|_{\e=0} \,.
\ee
We remind ourselves that the geometry $\hat B_n$ is the analytic continuation of the orbifold $B_n/Z_n$, and has a conical defect at a codimension 2 surface $C_n$ with deficit $2\pi\e=2\pi(1-1/n)$.

We emphasize again that $S[\hat B_n]$ does not include any contribution from the conical defect.  Therefore we may define it by excising a small region around $C_n$, say $\r<\r_0$, and calculate the action by integrating over $\r>\r_0$.  At the end of the calculation we may take the $\r_0\to0$ limit.

Before taking the $\r_0\to0$ limit, however, we have a boundary at $\r=\r_0$.  It is useful to close the boundary by filling in the excised region $\r<\r_0$ with a regular geometry.  This is called a regularized cone.

It is useful to consider the contribution to the action from the ``inside'' region $\r<\r_0$, even though this is not included in the definition of $S[\hat B_n]$.  Let us call it $S_\inside$.  From this perspective it is natural to rename $S[\hat B_n]$ as $S_\outside$.  Note that we have dropped the $\hat B_n$ dependence for notational simplicity.

The claim is that
\be\la{siso}
\lt. \pa_\e S_\outside \rt|_{\e=0} = -\lt. \pa_\e S_\inside \rt|_{\e=0} \,.
\ee
Therefore we instead of calculating \er{seeo} we may simply calculate
\be\la{seei}
S_{EE} = -\lt. \pa_\e S_\inside \rt|_{\e=0} \,,
\ee
which is a localized quantity and usually easier to calculate.

To show \er{siso} we note that the total action, $S_\total \eq S_\inside + S_\outside$, is defined on a manifold that has no boundary at $C_n$.  Furthermore it is regularized and not divergent.  These ensure that its variation from $\e=0$, a solution to the equations of motion, must vanish to linear order in $\e$.  From this we immediately find \er{siso}.  We show this pictorially in Figure~\re{figcone}.

\begin{figure}[h]
\centering
\includegraphics[width=0.8\textwidth]{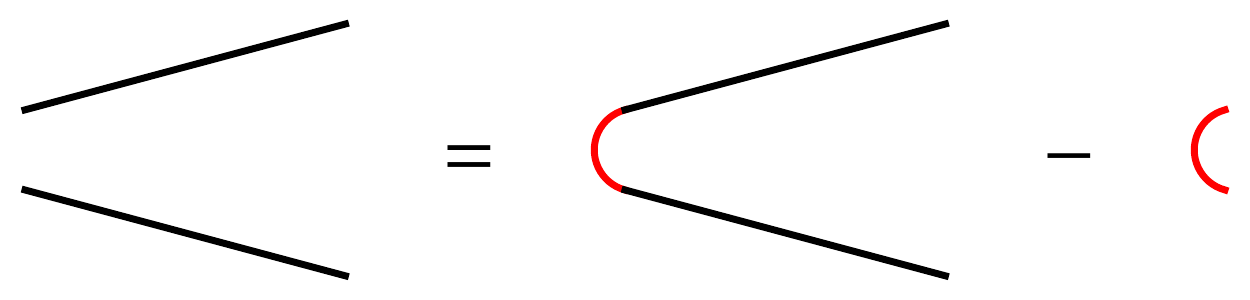}
\caption{\label{figcone}A pictorial way of filling in the tip of the cone and writing $S_\outside=S_\total-S_\inside$.}
\end{figure}

\sse{Derivation of the formula}\la{secee}

Let us now calculate $S_\inside$ to linear order in $\e$, for the regularized version of the conical configuration $\hat B_n$, in a general diffeomorphism-invariant Lagrangian built from contractions of an arbitrary number of Riemann tensors:
\be\la{act}
S=\int d^D x \sqrt{G} L(R_{\m\r\n\s}) \,.
\ee

A regularized cone may be written as
\be\la{met2}
ds^2 = e^{2A} (d\r^2 + \r^2 d\t^2) + (g_{ij} + 2K_{aij} x^a) dy^i dy^j + \cdots \,,
\ee
where the warp factor $A$ is regularized with a thickness parameter $a$ as
\be\la{areg}
A=-\fr\e2 \log(\r^2+a^2) \,.
\ee
This is the simplest choice for the regulator in our approach.  As we will see, the final result does not depend on the choice of the regulator because the coefficient of a would-be logarithmic divergence is universal (just like in an anomaly calculation).

There are two kinds of terms contributing linearly in $\e$.  Let us discuss them separately.

\ssse{First term (Wald's formula)}

The first contribution comes from a single Riemann tensor which gives
\be
R_{z\zb z\zb} = \fr14 e^{2A} \hat\na^2 A + \cdots
\ee
or the other three equivalent forms under the Riemann symmetry: $R_{z\zb\zb z}$, $R_{\zb zz\zb}$, and $R_{\zb z\zb z}$.  Here $z=\r e^{i\t}$, and the hat means covariant derivatives with respect to the metric with the singular warp factor stripped off: $\hat g_{ab} \eq e^{-2A} G_{ab}$.  The dots denote terms that are less singular and unimportant near $\r=0$.

To linear order in $\e$ this term contributes
\be
S_\inside^{(1)} = 4\int d^D x \sqrt{G} \lt. \fr{\pa L}{\pa R_{z\zb z\zb}} \rt|_{\e=0} \fr14 e^{2A} \hat\na^2 A \,,
\ee
where we have included a symmetry factor 4.  We define the derivative with respect to the Riemann tensor in the standard way, which is that the derivative has the Riemann symmetry and satisfies the identity $\d L = \fr{\pa L}{\pa R_{\m\r\n\s}} \d R_{\m\r\n\s}$.

Noting that $\hat\na^2 A$ is nothing but the (regularized) delta function $\d (x^1, x^2)$ times $-2\pi\e$, we may perform the integral along these 2 directions and find
\be
S_\inside^{(1)} = -2\pi\e \int d^d y \sqrt g \lt. \fr{\pa L}{\pa R_{z\zb z\zb}} \rt|_{\e=0}
\ee
to linear order in $\e$.  Using \er{seei} we find that the contribution to the entanglement entropy from this first kind of terms is
\be\la{ee1}
S_{EE}^{(1)} = 2\pi \int d^d y \sqrt g \fr{\pa L}{\pa R_{z\zb z\zb}} \,,
\ee
where we have dropped the $|_{\e=0}$.

This contribution is precisely Wald's formula for the black hole entropy in higher derivative gravity.  To see this we may easily covariantize \er{ee1} into
\be
S_{EE}^{(1)} = -2\pi \int d^d y \sqrt g \fr{\pa L}{\pa R_{\m\r\n\s}} \ve_{\m\r} \ve_{\n\s} \,,
\ee
which agrees with Wald's formula \er{wald}.  Here we have used the 2-dimensional Levi-Civita tensor in the $x^a$ directions.  In the complex coordinates it can be defined as $\ve_{z\zb}=-\ve_{\zb z}= iG_{z\zb}$, with all other components zero (including those along the $y^i$ directions).

\ssse{Second term (``anomaly'')}\la{secan}

There is a second way for a linear contribution to arise.  As we mentioned at the beginning of this section, it comes from the product of two Riemann tensors, one contributing the form
\be
R_{zizj} = 2 K_{zij} \na_z A + \cdots \,,
\ee
and the other contributing the conjugate form $R_{\zb k\zb l}$.

Note that this is the only way to get an $\mo(\e^2)$ contribution to the Lagrangian that may be enhanced to $\mo(\e)$ after the $\r$ integral.  For example, if both Riemann tensors contribute $\na_z A \approx -\e/(2z)$, we will be left with a factor $e^{-2i\t}$ under the $\t$ integral\footnote{This is correct at the order that we are considering, because at $\r=0$ and $\e=0$ the other Riemann tensors in $L$ all contribute terms that do not depend on $\t$.  These terms are worked out in \er{rexp}.}, which vanishes.  Furthermore, other components of the Riemann tensor, such as $R_{zi\zb j}$, $R_{abci}$, or $R_{abcd}$, do not contribute a term\footnote{This can be shown to be true for the more complete metric \er{met3}.} that scales like $1/\r$.

Therefore we should consider the following contribution to the action of the regularized cone:
\be\la{sin2}
S_\inside^{(2)} = 4^2 \int d^D x \sqrt{G} \fr{\pa^2 L}{\pa R_{zizj} \pa R_{\zb k\zb l}} (2 K_{zij} \na_z A) (2 K_{\zb kl} \na_\zb A) \,,
\ee
where we have included a symmetry factor $4^2$ to take into account equivalent forms such as $R_{zijz}$, $R_{izzj}$, and $R_{izjz}$ for $R_{zizj}$.

The $\r$ integral in \er{sin2} is of the form
\be
\int \r d\r (\na_z A) (\na_\zb A) e^{-\b A}
\ee
where we have included a factor $e^{-\b A}$ that comes from the rest of integrand.  In the regularized cone geometry this integral is simply
\ba
&\int_0^\infty \r d\r (\na_z A) (\na_\zb A) e^{-\b A} - (a\to0) \\
=\; &\fr{\e^2}{4} \int_0^\infty \r^3 d\r (\r^2+a^2)^{\fr{\b}{2}\e-2} - (a\to0) \,,
\ea
where we subtract off the contribution of the singular cone ($a=0$), and integrate from $\r=0$ to $\infty$.  This integral can be done exactly, giving
\be\la{int}
\fr{\e^2}{4} \[\lt. (\r^2+a^2)^{\fr{\b}{2}\e} \(\fr{1}{\b\e} - \fr{a^2}{(\r^2+a^2)(\b\e-2)}\) \rt|_0^\infty -(a\to0) \] \,.
\ee
As expected, the naively $\mo(\e^2)$ contribution is enhanced to $\mo(\e)$ because of the would-be logarithmic divergence at $\e=0$.  Following our regularization procedure, we should first keep the $\mo(\e)$ term in \er{int} and then take $a\to0$.  This gives
\be\la{rint}
\int \r d\r (\na_z A) (\na_\zb A) e^{-\b A} = -\fr{\e}{4\b} \,.
\ee

Even though we have used a particular regulator \er{areg}, we expect that this result is regulator-independent because it arises as the coefficient of a potential logarithmic divergence.

It remains to work out what $\b$ is.  There are 3 places where powers of $e^A$ might arise in \er{sin2}: from $\sqrt G$, the inverse metrics, and the Riemann tensors.  Let us write each term in $L$ as a contraction of Riemann tensors with all covariant (i.e.\ lower) indices together with twice as many inverse metrics.  In order to determine the powers of $e^A$ in each component of the Riemann tensor, we need to first keep all $\mo(\r^2)$ terms in the metric of the regularized cone:
\bm\la{met3}
ds^2 = e^{2A} \[dz d\zb + e^{2A} T (\zb dz-z d\zb)^2 \] + \(g_{ij} + 2K_{aij} x^a + Q_{abij} x^a x^b\) dy^i dy^j \\
+ 2i e^{2A} \(U_i + V_{ai} x^a\) \(\zb dz-z d\zb\) dy^i + \cdots \,.
\em
The derivation of this metric as well as the appearances of $e^{2A}$ is worked out in Appendix~\re{appsc}.  Note that we must in general allow off-diagonal components which may be concisely written as $G_{ai} = 2e^{2A} \(U_i + V_{ci} x^c\) \hat\ve_{ab} x^b$, where $\hat\ve_{ab}$ can be defined as $\hat\ve_{z\zb}=-\hat\ve_{\zb z}= i\hat g_{z\zb}$.  Before proceeding, we mention that all coefficient tensors in \er{met3} -- i.e.\ $T$, $g_{ij}$, $K_{aij}$, $Q_{abij}$, $U_i$, $V_{ai}$ -- when written in the $(z,\zb)$ basis are independent of $z$ or $\zb$ (although they could certainly depend on $y^i$), with the exception that $Q_{z\zb ij} = Q_{\zb zij}$ secretly\footnote{We could have made this manifest in the metric \er{met3} but that would make the equation longer.} contains a factor $e^{2A}$.  This is also derived in Appendix~\re{appsc}.

It is now straightforward to work out each component of the Riemann tensor and only keep terms that are important near $\r=0$:
\ba
R_{abcd} &= 12 e^{4A} T \hat\ve_{ab} \hat\ve_{cd} \,,\nn\\
R_{abci} &= 3 e^{2A} \hat\ve_{ab} V_{ci} \,,\nn\\
R_{abij} &= 2 e^{2A} \hat\ve_{ab} (\pa_i U_j - \pa_j U_i) + g^{kl} (K_{ajk} K_{bil} - K_{aik} K_{bjl}) \,,\la{rexp}\\
R_{aibj} &= e^{2A} \[ \hat\ve_{ab} (\pa_i U_j - \pa_j U_i) + 4\hat g_{ab} U_i U_j \] + g^{kl} K_{ajk} K_{bil} - Q_{abij} \,,\nn\\
R_{aijk} &= \pa_k K_{aij} - \T\g{l}{ik} K_{ajl} + 2 \hat\ve_{ab} \hat g^{bc} K_{cij} U_k - (j\lra k)\,,\nn\\
R_{ikjl} &= r_{ikjl} + e^{-2A} \hat g^{ab} (K_{ail} K_{bjk} - K_{aij} K_{bkl}) \,,\nn
\ea
where we have used $\T \g{l}{jk}$ and $r_{ikjl}$ to denote the intrinsic Christoffel symbol and Riemann tensor for $g_{ij}$ respectively.

For each component of the Riemann tensor shown above we should count the power of $e^A$.  We may simplify our task by combining the powers of $e^A$ from each Riemann component with those from the associated inverse metrics.  This means that we associate an extra factor $e^{-4A}$ to $R_{abcd}$, an extra factor $e^{-3A}$ to $R_{abci}$, etc.  In other words, we push possible powers of $e^A$ from the inverse metrics into the Riemann tensors by going to the proper coordinates in the $x^a$ directions.

Combining the above considerations, we find that we should associate $e^{-2A}$ to $Q_{zzij}$ and $Q_{\zb\zb ij}$, as well as $e^{-A}$ to $V_{ai}$ and $K_{aij}$, and we should associate no power of $e^A$ to the other terms in the expansion ($T$, $U_i$, $Q_{z\zb ij}$, and $r_{ikjl}$).

Let us therefore fully expand $\fr{\pa^2 L}{\pa R_{zizj} \pa R_{\zb k\zb l}}$ and denote each term in the expansion as $(\fr{\pa^2 L}{\pa R_{zizj} \pa R_{\zb k\zb l}})_\a$, where $\a$ runs over all the terms.  Each term is a product of quantities appearing on the right hand sides of \er{rexp}.  Let us use $q_\a$ to count the total number of $Q_{zzij}$ and $Q_{\zb\zb ij}$, plus one half of the total number of $V_{ai}$ and $K_{aij}$ in each term.  Then for such a term in the integrand of \er{sin2} we have in total
\be
e^{2A} e^{-4A} e^{-2q_\a A} = e^{-2(q_\a+1)A} \,,
\ee
where $e^{2A}$ comes from $\sqrt G$, and $e^{-4A}$ comes from the two powers of $G^{z\zb}$ that we have not yet counted because they are associated with the $R_{zizj} R_{\zb k\zb l}$ which we stripped off by taking the second derivative of $L$.  Note that once we have counted the powers of $e^A$ we may safely set $A=0$.

Plugging $\b=2(q_\a+1)$ into \er{rint} and \er{sin2}, we find
\be
S_\inside^{(2)} = -16\pi\e \int d^d y \sqrt{g} \sum_\a \(\fr{\pa^2 L}{\pa R_{zizj} \pa R_{\zb k\zb l}}\)_\a \fr{K_{zij} K_{\zb kl}}{q_\a+1} \,,
\ee
where a sum over $\a$ is implied.  Therefore the second kind of contribution to the entanglement entropy is
\be\la{ee2}
S_{EE}^{(2)} = 16\pi \int d^d y \sqrt{g} \sum_\a \(\fr{\pa^2 L}{\pa R_{zizj} \pa R_{\zb k\zb l}}\)_\a \fr{K_{zij} K_{\zb kl}}{q_\a+1} \,.
\ee

Note that we do not actually have to expand $R_{abcd}$, $R_{abci}$, and $R_{aijk}$, because they all consist of terms that have the same power of $e^A$.  Therefore, we may evaluate \er{ee2} simply as follows.  We expand only the following components of the Riemann tensor
\ba
R_{abij} &= \td R_{abij} + g^{kl} (K_{ajk} K_{bil} - K_{aik} K_{bjl}) \,,\nn\\
R_{aibj} &= \td R_{aibj} + g^{kl} K_{ajk} K_{bil} - Q_{abij} \,,\la{rexps}\\
R_{ikjl} &= r_{ikjl} + \hat g^{ab} (K_{ail} K_{bjk} - K_{aij} K_{bkl}) \,,\nn
\ea
where we have defined the shorthand notation\footnote{We have set $A=0$ in both \er{rexps} and \er{rtd} as we have already counted the powers of $e^A$.}
\ba
\td R_{abij} &\eq 2 e^{2A} \hat\ve_{ab} (\pa_i U_j - \pa_j U_i) \,,\nn\\
\td R_{aibj} &\eq e^{2A} \[ \hat\ve_{ab} (\pa_i U_j - \pa_j U_i) + 4\hat g_{ab} U_i U_j \] \,.\la{rtd}
\ea
Once we have expanded the second of derivative of $L$ according to \er{rexps}, for each term (which is a product) we define $q_\a$ as the total number of $Q_{zzij}$ and $Q_{\zb\zb ij}$, plus one half of the total number of $K_{aij}$, $R_{abci}$, and $R_{aijk}$.\footnote{We thank Joan Camps for pointing out the omission of $R_{abci}$ in a previous version of this paper.}  Finally we sum over $\a$ with weights $1/(1+q_\a)$.  We can then eliminate $\td R_{abij}$, $\td R_{aibj}$, and $r_{ikjl}$ (if we want) using \er{rexps}, arriving at an expression involving only components of $R_{\m\n\r\s}$, $K_{aij}$ and $Q_{abij}$.  Therefore we do not actually need to know the definitions of $\td R_{abij}$ and $\td R_{aibj}$.  The expansion and resummation can be thought of as a simple prescription to generate higher order terms in $K_{aij}$ and $Q_{abij}$.

Before proceeding, let us comment on the similarity of this calculation with the calculation of the Weyl anomaly.  In the latter, we frequently separate the regulated effective action $W$ into a renormalized effective action $W_{\rm fin}$ and a counterterm $W_{\rm ct}$.  Since $W$ is invariant under a Weyl transformation combined with a compensating scaling of the cutoff, we conclude that the variation of $W_{\rm fin}$ must be minus that of $W_{\rm ct}$.  Here we may draw an analogy between $S_\outside$ and $W_{\rm fin}$, as well as between $S_\inside$ with $W_{\rm ct}$.  Therefore in a certain sense, the extrinsic curvature corrections \er{ee2} to Wald's formula may be thought of as ``anomalies,'' and the coefficients $q_\a$ may be thought of as ``anomaly coefficients.''

Finally, let us combine \er{ee1} and \er{ee2}, arriving at our entropy formula
\be\la{ee}
S_{EE}= 2\pi\int d^d y \sqrt{g} \lt\{ \fr{\pa L}{\pa R_{z\zb z\zb}} + \sum_\a \(\fr{\pa^2 L}{\pa R_{zizj} \pa R_{\zb k\zb l}}\)_\a \fr{8K_{zij} K_{\zb kl}}{q_\a+1} \rt\} \,.
\ee
This formula, when evaluated on $C_1$, gives the holographic entanglement entropy in a general theory dual to higher derivative gravity.  We emphasize that $C_1$ is the $n\to1$ limit of the conical defect $C_n$, whose location can in principle be determined by solving all bulk equations of motion with conical boundary conditions.  For the prescription on how to calculate the second term in \er{ee} we refer our readers to the paragraphs below \er{eei} or \er{ee2}.

Before continuing, let us point out that this formula \er{ee} can also be used for the black hole entropy, in which case it is evaluated on the horizon.  In particular, if the horizon is a Killing horizon, the extrinsic curvature vanishes and we recover Wald's formula.  This can be thought of as a holographic derivation of Wald's formula (for Killing horizons) in general higher derivative gravity.

Finally, let us comment on possible power law divergences in the action $S[\hat B_n]$.  They arise from more than one factor of $K_{zizj} K_{\zb k\zb l}$ contributing to the action.  Each such factor (together with two corresponding powers of $G^{z\zb}$) contributes $\e^2 \r^{4\e-2}$ to the $\r$ integral, resulting in power-law divergences for small $\e$.  However, this is an artifact of analytic continuation, as we can see by going back to the physical cases with integer $n\ge2$.  In these cases, we have $\e\ge1/2$, and factors of $\r^{4\e-2}$ do not cause any divergences.  In principle we should then analytically continue the result of the integral to $\e<1/2$.  This is difficult in practice.  Therefore we choose to analytically continue the integrand to $\e<1/2$, and we need to introduce local counterterms in $S_\inside$ to cancel possible divergences (which are simply artifacts of analytic continuation).  This is analogous to counterterms and renormalization in field theories.  Here, we happily find that power-law divergences do not affect our result because they come with even more powers of $\e$ than logarithmic divergences.

\ssse{Covariant form}

If desired, we may transform \er{ee} into a covariant form
\bm\la{eecov}
\!\!\!\!\!
S_{EE}= 2\pi\int d^d y \sqrt{g} \lt\{ -\fr{\pa L}{\pa R_{\m\r\n\s}} \ve_{\m\r} \ve_{\n\s} + \sum_\a \(\fr{\pa^2 L}{\pa R_{\m_1\r_1\n_1\s_1} \pa R_{\m_2\r_2\n_2\s_2}}\)_\a \fr{2K_{\l_1\r_1\s_1} K_{\l_2\r_2\s_2}}{q_\a+1} \times \rt. \\
\lt.\phantom{\frac12} \times \[ (n_{\m_1\m_2} n_{\n_1\n_2}-\ve_{\m_1\m_2} \ve_{\n_1\n_2}) n^{\l_1\l_2} + (n_{\m_1\m_2} \ve_{\n_1\n_2}+\ve_{\m_1\m_2} n_{\n_1\n_2}) \ve^{\l_1\l_2}\] \rt\} \,.
\em
Here the second line is simply a projector which enforces $\m_1=\n_1=\l_1 \ne \m_2=\n_2=\l_2$ in the $(z,\zb)$ basis.  We have used $n_{\m\n}$ which projects to the induced 2-dimensional metric $G_{ab}$ in the $x^a$ directions.  In terms of two orthogonal unit vectors $n^{(a)}_\m$ in these directions (which by an abuse of notation is also called $n$) we may define it as
\be
n_{\m\n} = n^{(a)}_\m n^{(b)}_\n G_{ab} \,.
\ee
The $\ve_{\m\n}$ tensor is defined as before and can now be written as
\be
\ve_{\m\n} = n^{(a)}_\m n^{(b)}_\n \ve_{ab} \,,
\ee
where $\ve_{ab}$ is the usual Levi-Civita tensor.  The product of two $\ve$ tensors may be rewritten as
\be
\ve_{\m\n} \ve_{\r\s} = n_{\m\r} n_{\n\s} - n_{\m\s} n_{\n\r} \,.
\ee
We can similarly define $K_{\l\m\n}$ in terms of the usual extrinsic curvature $K_{aij}$:
\be
K_{\l\m\n} = n^{(a)}_\l m^{(i)}_\m m^{(j)}_\n K_{aij} \,,
\ee
where $m^{(i)}_\m$ are a set of $d$ orthogonal unit vectors along the $y^i$ directions.

\sse{Examples}\la{secex}

\ssse{f(R) gravity}
In $f(R)$ gravity, the Lagrangian $L$ depends on the Riemann tensor only through the Ricci scalar $R$.  Using our general formula \er{ee}, we arrive at
\be\la{eefr}
S_{EE}= -4\pi\int d^d y \sqrt{g} \fr{\pa L}{\pa R} \,.
\ee
The second term in \er{ee} involving extrinsic curvatures vanishes because $R$ does not contain components of the form $R_{zizj}$.

This is a (trivial) consistency check on our formula because we may transform $f(R)$ gravity to a theory of Einstein gravity coupled to a scalar, and using the Ryu-Takayanagi formula we find the same entanglement entropy \er{eefr}.

\ssse{General four-derivative gravity}
At four derivatives\footnote{By this we mean four derivatives (from two Riemann tensors) in the Lagrangian.  This seems to be the convention in the literature.}, the most general Lagrangian involving only the Riemann tensor is
\be\la{L4der}
L = \l_1 R^2 + \l_2 R_{\m\n} R^{\m\n} + \l_3 R_{\m\r\n\s} R^{\m\r\n\s} \,.
\ee
If we prefer we can always add an Einstein-Hilbert term (and/or a cosmological constant), which simply adds an area term to the entanglement entropy.

Focusing on the nontrivial part \er{L4der}, we use our general formula \er{ee} and after a straightforward calculation we find
\be
S_{EE}= -4\pi\int d^d y \sqrt{g} \[ 2\l_1 R + \l_2 \(\T Raa -\fr12 K_a K^a\) + 2\l_3 \(\T R {ab}{ab} - K_{aij} K^{aij}\) \]
\ee
where we have covariantized the answer and used the trace of the extrinsic tensor $K_a \eq K_{aij}g^{ij}$.  Note that the calculation is particularly simple because all $q_\a=0$.

This reproduces the result reached in \cite{Fursaev:2013fta} by an explicit calculation using a different regulator.  It also agrees with an independent calculation in \cite{Myers:2013lva} using the Randall-Sundrum II braneworld.  This result has been used to calculate the holographic entanglement entropy in particular theories of four-derivative gravity in e.g.\ \cite{Bhattacharyya:2013gra, Alishahiha:2013zta}.

\ssse{Lovelock gravity}\la{seceell}
Let us consider Lovelock gravity at $2p$ derivatives, keeping $p$ arbitrary.  The Lagrangian can be conveniently written as
\be\la{lll}
L_{2p}(R) = \fr{1}{2^p} \d^{\m_1\r_1\m_2\r_2\cdots\m_p\r_p}_{\n_1\s_1\n_2\s_2\cdots\n_p\s_p} \Tu R {\m_1\r_1}{\n_1\s_1} \Tu R {\m_2\r_2}{\n_2\s_2} \cdots \Tu R {\m_p\r_p}{\n_p\s_p}
\ee
up to an overall coupling constant.  Here the generalized delta symbol is defined as the totally antisymmetric product of $2p$ Kronecker delta symbols, or defined recursively as
\be
\d^{\m_1\m_2\cdots\m_n}_{\n_1\n_2\cdots\n_n} = \sum_{i=1}^n (-1)^{i+1} \d^{\m_1}_{\n_i} \d^{\m_2\m_3\cdots\m_n}_{\n_1\cdots\hat\n_i\cdots\n_n} \,.
\ee

Let us derive the entanglement entropy using the general formula \er{ee}.  The first term is
\be\la{ll1}
\fr{\pa L_{2p}(R)}{\pa R_{z\zb z\zb}} = \fr{p}{2^p} \d^{\m_1\r_1\cdots\m_{p-1}\r_{p-1}z\zb}_{\n_1\s_1\cdots\n_{p-1}\s_{p-1}\zb z} \Tu R {\m_1\r_1}{\n_1\s_1} \cdots \Tu R {\m_{p-1}\r_{p-1}}{\n_{p-1}\s_{p-1}} (G^{z\zb})^2 \,.
\ee
Since the generalized delta symbol is totally antisymmetric in all upper indices as well as in all lower indices, we conclude that the dummy indices in \er{ll1} can neither be $z$ nor $\zb$, and must therefore be along the $y^i$ directions.  Therefore \er{ll1} reduces to
\be\la{ll12}
\fr{\pa L_{2p}(R)}{\pa R_{z\zb z\zb}} = - \fr{p}{2^{p-2}} \d^{i_1k_1\cdots i_{p-1}k_{p-1}}_{j_1l_1\cdots j_{p-1}l_{p-1}} \Tu R {i_1k_1}{j_1l_1} \cdots \Tu R {i_{p-1}k_{p-1}}{j_{p-1}l_{p-1}} \,.
\ee
By the same argument the second term in \er{ee} becomes
\bm\la{ll2}
\fr{\pa^2 L_{2p}(R)}{\pa R_{zizj} \pa R_{\zb k\zb l}} 8K_{zij} K_{\zb kl} \\
= - \fr{p}{2^{p-2}} 8(p-1) \d^{i_1k_1\cdots i_{p-1}k_{p-1}}_{j_1l_1\cdots j_{p-1}l_{p-1}} \Tu R {i_1k_1}{j_1l_1} \cdots \Tu R {i_{p-2}k_{p-2}}{j_{p-2}l_{p-2}} \Tu K {zi_{p-1}}{j_{p-1}} \Tu K {\zb k_{p-1}}{l_{p-1}} \,,
\em
where we have not yet included the factor $1/(q_\a+1)$.

We claim that the total of these two contributions \er{ll12} and \er{ll2} is
\ba\la{eels}
S_{EE} &= -4\pi p \int d^d y \sqrt{g} L_{2p-2}(r) \\
&= -2\pi \int d^d y \sqrt{g} \fr{p}{2^{p-2}} \d^{i_1k_1\cdots i_{p-1}k_{p-1}}_{j_1l_1\cdots j_{p-1}l_{p-1}} \Tu r {i_1k_1}{j_1l_1} \cdots \Tu r {i_{p-1}k_{p-1}}{j_{p-1}l_{p-1}} \,,\la{eell}
\ea
where as before $\Tu r {ik}{jl}$ denotes the intrinsic Riemann tensor of the conical defect.  This agrees exactly with the Jacobson-Myers prescription for the black hole entropy in Lovelock gravity \cite{Jacobson:1993xs}, which differs from Wald's formula by precisely the ``anomaly''-like terms \er{ll2}.  The Jacobson-Myers formula used as the holographic entanglement entropy for Gauss-Bonnet gravity leads to the correct universal terms expected for a generic 4-dimensional CFT \cite{Hung:2011xb}, including extrinsic curvature terms first worked out in \cite{Solodukhin:2008dh}.  Other discussions on holographic entanglement entropy in Lovelock gravity include \cite{deBoer:2011wk, Safdi:2012sn}.  A recent proposal \cite{Sarkar:2013swa} for the gravitational entropy at the quadratic order in $K$ in f(Lovelock) gravity can be shown to agree with our general formula \er{ee} by the same argument presented here. 

Let us now prove \er{eell}.  First, we simplify the notation by defining
\be
u_s^{(0)} \eq \Tu R {i_sk_s}{j_sl_s} \,,\qu
u_s^{(1)} \eq \Tu r {i_sk_s}{j_sl_s} \,,\qu
u_s^{(2)} \eq \Tu K {ai_s}{l_s} \T K {aj_s}{k_s} - \Tu K {ai_s}{j_s} \T K {al_s}{k_s} \,,
\ee
where $s=1,2,\cdots,p-1$.  Then from the last equation in \er{rexps} we have
\be
u_s^{(0)} = u_s^{(1)} + u_s^{(2)} \,.
\ee
Using the antisymmetry of the generalized delta symbol, we may make the following replacement in \er{ll2}:
\be
\Tu K {zi_{p-1}}{j_{p-1}} \Tu K {\zb k_{p-1}}{l_{p-1}} \to -\fr18 u_{p-1}^{(2)} \,,
\ee
and furthermore replace
\be
(p-1) u_1^{(1)} \cdots u_{p-2}^{(1)} u_{p-1}^{(2)} \to \sum_{s=1}^{p-1} u_1^{(1)} \cdots \hat u_s^{(1)} u_s^{(2)} \cdots u_{p-1}^{(1)} \,,
\ee
where the summand is the product of all $u^{(1)}$ except that $u_s^{(1)}$ is replaced by $u_s^{(2)}$.  Then all that we need in order to prove \er{eell} is
\be\la{llid}
u_1^{(0)} \cdots u_{p-1}^{(0)} - \sum_{s=1}^{p-1} \fr{1}{q_\a+1} \(u_1^{(1)} \cdots \hat u_s^{(1)} u_s^{(2)} \cdots u_{p-1}^{(1)}\)_\a = u_1^{(2)} \cdots u_{p-1}^{(2)} \,.
\ee
To prove this, we expand the left hand side using $u_s^{(0)} = u_s^{(1)} + u_s^{(2)}$.  Recall that $q_\a$ counts the number of $u^{(2)}$ in the expansion, not counting $u_s^{(2)}$ since it does not come from the second derivative of $L$.  Therefore $q_\a+1$ simply counts the total number of $u^{(2)}$ in the expansion including $u_s^{(2)}$.

Each term in the expansion of $u_1^{(0)} \cdots u_{p-1}^{(0)}$ can be written as $u_1^{(t_1)} \cdots u_{p-1}^{(t_{p-1})}$ where $t_1, \cdots, t_{p-1}$ take values of either 1 or 2.  The same term appears in the sum of \er{llid}, each time with weight $1/(q_\a+1)$.  The total number of times that it appears is precisely the number of $u^{(2)}$, i.e.\ $(q_\a+1)$, so the terms are canceled on the left hand side of \er{llid}.  This is true except for the term $u_1^{(1)} \cdots u_{p-1}^{(1)}$, which is the right hand side.

This completes our proof of the entanglement entropy formula \er{eels} in general Lovelock gravity.

\se{Minimization of the entropy formula}\la{secmin}

In Section~\re{secee} we derived a formula \er{ee} that, when evaluated on $C_1$, gives the holographic entanglement entropy in a general theory dual to higher derivative gravity.

The surface $C_1$ is well-defined -- it is the $n\to1$ limit of the conical defect $C_n$, whose location can in principle be determined by solving all bulk equations of motion with conical boundary conditions.  This may be difficult in practice, and an alternative method of finding the location of $C_1$ is desirable.

A natural conjecture is that it is determined by minimizing the same entropy formula \er{ee}, analogous to the Ryu-Takayanagi prescription in Einstein gravity.  In this section, we will show that this is true in the three examples considered so far: $f(R)$, general four-derivative, and Lovelock gravity.  We leave a general solution to this problem for future work.

\sse{Lovelock gravity}

Before discussing the problem in general, it is useful to solve it first in the case of Lovelock gravity.  The reason is two-fold: first, the formalism of Lovelock gravity is particularly simple and clear; and second, there is a subtlety in the derivation that already appears in Lovelock gravity \cite{Bhattacharyya:2013jma, Chen:2013qma, Bhattacharyya:2013gra}.

Recall that in the derivation of the Ryu-Takayagani prescription, we introduced two equivalent methods of analytically continuing the conical geometry $\hat B_n$ and finding the location of $C_n$ as $n\to1$: the boundary condition method and the cosmic brane method.

We will see how each method works for general Lovelock gravity.

\ssse{Boundary condition method}

Recall that in the boundary condition method of Section~\re{sectm}, we found the location of $C_1$ by looking at a potentially divergent term in the $zz$ component of the Einstein equation.  Let us follow the same strategy.  The derivation here is a straightforward generalization of the discussion in \ci{Bhattacharyya:2013gra} for Gauss-Bonnet to general Lovelock gravity.

The equation of motion for Lovelock gravity at $2p$ derivatives \er{lll} can be nicely written as
\be\la{lleom}
\T E {\m}{\n} = \fr{1}{2^{p+1}} \d^{\m\m_1\r_1\m_2\r_2\cdots\m_p\r_p}_{\n\n_1\s_1\n_2\s_2\cdots\n_p\s_p} \Tu R {\m_1\r_1}{\n_1\s_1} \Tu R {\m_2\r_2}{\n_2\s_2} \cdots \Tu R {\m_p\r_p}{\n_p\s_p} \,,
\ee
where $E^{\m\n} \eq \fr{1}{\sqrt G} \fr{\d S}{\d G_{\m\n}}$.

Let us look at $E_{zz}$, so in \er{lleom} we set $\m=\zb$, $\n=z$.  Due to the antisymmetry of the generalized delta symbol, there is at most one $z$ in the upper indices and at most one $\zb$ in the lower indices.  In this case we can always consider $\m_1=z$, $\n_1=\zb$, and include a symmetry factor $(2p)^2$.  All other dummy indices in \er{lleom} must be along the $y^i$ directions.  Therefore, the equation of motion becomes
\ba
\T E {\zb}{z} &= \fr{(2p)^2}{2^{p+1}} \d^{\zb z\r_1\m_2\r_2\cdots\m_p\r_p}_{z\zb\s_1\n_2\s_2\cdots\n_p\s_p} \Tu R {z\r_1}{\zb\s_1} \Tu R {\m_2\r_2}{\n_2\s_2} \cdots \Tu R {\m_p\r_p}{\n_p\s_p} +\cdots \\
&= -\fr{p^2}{2^{p-2}} \fr{\e}{z} \Tu K {zi}{j}\d^{ii_1k_1\cdots i_{p-1}k_{p-1}}_{jj_1l_1\cdots j_{p-1}l_{p-1}} \Tu R {i_1k_1}{j_1l_1} \cdots \Tu R {i_{p-1}k_{p-1}}{j_{p-1}l_{p-1}} +\cdots \,,\la{llezz}
\ea
where we have used $R_{zizj}= \e K_{zij}/z +\cdots$, and the dots denote terms that are less divergent as $\r\to0$.  This includes terms from \er{lleom} in which none of the dummy indices are $z$ or $\zb$.

Again we argue that the potential $1/z$ divergence must vanish because the stress tensor from the matter sector (if present) is not expected to diverge.  There is, however, a subtlety here.  The Riemann tensors in \er{llezz} are all projected to the $y^i$ directions.  If we decompose them according to the last equation in \er{rexp}:
\be\la{gc}
\Tu R{ik}{jl}=\Tu r{ik}{jl} + e^{-2A} \hat g^{ab} (\Tu K{ai}{l} \Tu K{bj}{k} - \Tu K{ai}{j} \Tu K{bk}{l}) \,,
\ee
we find that each pair of $K$ comes with a factor $e^{-2A}=\r^{2\e}$.  For finite $\e>0$, terms involving one or more pairs of $K$ are less divergent than the term
\be\la{ller}
\T E {\zb}{z} = -\fr{p^2}{2^{p-2}} \fr{\e}{z} \Tu K {zi}{j}\d^{ii_1k_1\cdots i_{p-1}k_{p-1}}_{jj_1l_1\cdots j_{p-1}l_{p-1}} \Tu r {i_1k_1}{j_1l_1} \cdots \Tu r {i_{p-1}k_{p-1}}{j_{p-1}l_{p-1}} +\cdots \,.
\ee
Therefore, at small but finite $\e$, we should set the coefficient of the most divergent term \er{ller} to zero\footnote{This is the only contribution to the most divergent term in the equation of motion if the conical metric has the form \er{met3} for small but finite $\e$, as derived in Appendix~\re{appsc}.}.  This gives
\be\la{llk}
\Tu K {zi}{j}\d^{ii_1k_1\cdots i_{p-1}k_{p-1}}_{jj_1l_1\cdots j_{p-1}l_{p-1}} \Tu r {i_1k_1}{j_1l_1} \cdots \Tu r {i_{p-1}k_{p-1}}{j_{p-1}l_{p-1}} =0 \,.
\ee
This equation can then be used in the $\e\to0$ limit to fix the location of $C_1$.  It is the analog of the $K_z=0$ condition \er{kz} in the case of Einstein gravity.

It is now straightforward to show that the same equation \er{llk} arises from minimizing the entanglement entropy \er{eels} for Lovelock gravity.  Since $S_{EE}$ depends only on induced metric $g_{ij}$ (and intrinsic curvature) in $d$ dimensions, the equation of motion for the embedding coordinates can be easily worked out:
\be\la{llp}
\Pi_a = 2 K_{aij} \fr{\d S_{EE}}{\d g_{ij}} =0 \,.
\ee
This was derived in e.g.\ \cite{Bhattacharyya:2013jma}.  Note that this is $K_{aij}$ contracted with the equation of motion for $S_{EE}$ which is simply another Lovelock theory at $2p-2$ derivatives.  Its equation of motion is analogous to \er{lleom}, and therefore we conclude that \er{llp} is exactly the same as \er{llk} (and its conjugate).

\ssse{Cosmic brane method}\la{seclb}

It is just as easy to show that minimizing $S_{EE}$ gives $C_1$ in Lovelock gravity using the cosmic brane method.  This amounts to showing that $S_B\eq \e S_{EE}$ serves as the action of a cosmic brane that creates a conical deficit $2\pi\e$, at least to linear order in $\e$.  

We do this by matching the delta functions in the equation of motion.  First, the part of the equation from the brane action $S_B\eq \e S_{EE}$ is
\be\la{llte}
\T {\td E} {i}{j} = -2p\tdd \fr{1}{2^p} \d^{ii_1k_1\cdots i_{p-1}k_{p-1}}_{jj_1l_1\cdots j_{p-1}l_{p-1}} \Tu r {i_1k_1}{j_1l_1} \cdots \Tu r {i_{p-1}k_{p-1}}{j_{p-1}l_{p-1}} \,,
\ee
where $\td E^{\m\n} \eq \fr{1}{\sqrt G} \fr{\d S_B}{\d G_{\m\n}}$, and we have introduced the shorthand notation
\be\la{tdd}
\tdd \eq 2\pi\e \fr{\d^{(2)}(x^1, x^2)}{\sqrt{g^{(2)}}} \,,
\ee
which will be used frequently in the rest of the paper.  Note that $g^{(2)}$ is the determinant of the induced metric in the $x^a$ directions.  All other components of $\td E$ vanish.

The part of the equation of motion from the Lovelock action is \er{lleom} which we repeat here:
\be\la{lle2}
\T E {\m}{\n} = \fr{1}{2^{p+1}} \d^{\m\m_1\r_1\m_2\r_2\cdots\m_p\r_p}_{\n\n_1\s_1\n_2\s_2\cdots\n_p\s_p} \Tu R {\m_1\r_1}{\n_1\s_1} \Tu R {\m_2\r_2}{\n_2\s_2} \cdots \Tu R {\m_p\r_p}{\n_p\s_p} \,.
\ee
Our goal is to argue that to solve the total equation of motion, we need to have $A=-\e\log\r$ near $\r=0$ which creates the desired conical deficit.  Here $A$ is the warp factor appearing in the metric \er{met3}.

Let us match the delta functions.  The only way to generate a delta function in \er{lle2} is when one of Riemann tensors contributes
\be\la{rz}
\Tu R {z\zb}{z\zb} = -e^{-2A} \hat\na^2 A +\cdots = \tdd + \cdots \,.
\ee
Happily, to linear order in $\e$ we never get two or more delta functions.  Let us take the first Riemann tensor to be \er{rz}, and include a symmetry factor $4p$ because there are $p$ Riemann tensors and 4 equivalent forms of \er{rz}.  Using the antisymmetry of the generalized delta symbol, all other indices must be along the $y^i$ directions.  This gives
\be\la{lle3}
\T E {i}{j} = 2p\tdd \fr{1}{2^p} \d^{ii_1k_1\cdots i_{p-1}k_{p-1}}_{jj_1l_1\cdots j_{p-1}l_{p-1}} \Tu R {i_1k_1}{j_1l_1} \cdots \Tu R {i_{p-1}k_{p-1}}{j_{p-1}l_{p-1}} \,,
\ee
which is precisely minus \er{llte} except for the difference between the projected and intrinsic Riemann tensors.  As we argued around \er{gc}, the difference is suppressed by factors of $\r^{2\e}$.  For finite $\e>0$, these terms vanish at the delta function.  More precisely, since the variation of the action is of the form
\be
\d S = \int d^Dx \sqrt G E^{\m\n} \d G_{\m\n} \,,
\ee
terms from \er{lle3} that have powers of $\r^{2\e}$ vanish under the integral.  We should evaluate them at finite $\e$ first, and then take the $\e\to0$ limit, as we are doing an analytic continuation.  This completes our derivation that $S_{EE}$ serves as the cosmic brane action in Lovelock gravity, and therefore minimizing it gives the location of $C_1$.

\sse{General case}

In this subsection, we work towards a general derivation of the location of $C_1$ in terms of minimization of some functional.  We will not be able to prove the conjecture that minimizing our entropy formula always gives $C_1$, but will show that this is true in all three examples discussed in Section~\re{secex}.

We will keep the derivation as general as possible, so that it works for the three examples simultaneously.  We feel that this could help us build a full proof (or disproof) of the conjecture in the future.

Let us use the cosmic brane method.  We would like to derive the cosmic brane action that creates a conical deficit $2\pi\e$, to linear order in $\e$.  As before, we do this by matching the delta functions in the equation of motion.  One difference from the Lovelock case discussed in Section~\re{seclb} is that in general we have derivatives acting on Riemann tensors, producing derivatives of a delta function.  Our prescription is therefore to match all derivatives of the delta function with respect to $x^a$, as well as delta functions with no derivatives if they do not come with any factors of $\r^{2\e}$.

For this discussion it is useful to consider the Lagrangian $L$ as a function of $\T R{\m}{\r\n\s}$ and $G_{\l\h}$.  This is because the variation of the Riemann tensor in this form is simply a total derivative:
\be
\d \T R{\m}{\r\n\s} = -2 \na^\m \na_\n \d G_{\r\s} |_{\sym(\m\r\n\s)} \,,
\ee
where by $|_{\sym(\m\r\n\s)}$ we mean that the equation holds after fully contracting it with a tensor that has the Riemann symmetry.

The part of the equation of motion from the Lagrangian $L$ is
\be\la{eom}
E^{\r\s} \eq \fr{1}{\sqrt G} \fr{\d S}{\d G_{\r\s}} = \fr12 G^{\r\s} L + \fr{\pa L}{\pa G_{\r\s}} - (\na^\m \na_\n + \na_\n \na^\m) \fr{\pa L}{\pa \T R{\m}{\r\n\s}} \,,
\ee
where we emphasize that $\pa L/\pa G_{\r\s}$ should be taken with $\T R{\m}{\r\n\s}$ fixed.

There are two ways for a delta function to arise in \er{eom}, which we will compensate by the equation of motion from two separate terms of the brane action.

\ssse{First term of the brane action}

First, one of the Riemann tensors in $L$ can simply contribute
\be
R_{z\zb z\zb} = \fr14 e^{2A} \hat\na^2 A + \cdots = -(G_{z\zb})^2 \tdd + \cdots \,,
\ee
where $\tdd$ is defined in \er{tdd} as before.  This gives the following contribution to the equation of motion \er{eom}:
\be\la{eom1}
E_{(1)}^{\r\s} = -4 \[ \fr12 G^{\r\s} + \fr{\pa}{\pa G_{\r\s}} - (\na_\m \na_\n + \na_\n \na_\m) \fr{\pa}{\pa R_{\m\r\n\s}} \] \fr{\pa L}{\pa R_{z\zb z\zb}} (G_{z\zb})^2 \tdd \,,
\ee
where the derivatives act all the way to the right, and we have included a symmetry factor 4 as before.  This contribution contains delta functions with 0, 1, or 2 derivatives.

It is easy to see that \er{eom1} is compensated by the equation of motion from the following brane action:
\be\la{b1}
S_B^{(1)} = 4 \int d^D x \sqrt G \fr{\pa L}{\pa R_{z\zb z\zb}} (G_{z\zb})^2 \tdd \,.
\ee
To linear order in $\e$, this is simply $\e$ times the first term \er{ee1} in our entropy formula, i.e.\ Wald's formula.

\ssse{Second term of the brane action}

The second way of getting a delta function in \er{eom} is when one of the Riemann tensors in $\pa L/\pa \T R{\m}{\r\n\s}$ contributes $\na_a A$ or powers of $e^{-2A}$, as the two derivatives $\na^\m \na_\n$ potentially act on them and produce a delta function or its first derivative.

Let us first focus on terms with the derivative of the delta function, which arise only if $\pa L/\pa \T R{\m}{\r\n\s}$ has a term of the form
\be
R_{zizj} = 2 K_{zij} \na_z A + \cdots \,,
\ee
or its conjugate $R_{\zb i\zb j}$.  This contributes to \er{eom} as
\be\la{eom2}
E_{(2)}^{\r\s} = -16 \na_\m \na_\n \( \fr{\pa^2 L}{\pa R_{zizj} \pa R_{\m\r\n\s}} K_{zij} \na_z A\) + (z\to \zb)\,,
\ee
where we have included a symmetry factor 4.  The only way for the derivative of the delta function to arise is when either $\m$ or $\n$ is $\zb$, and the other one is $z$ or $\zb$.  Then both derivatives act on $\na_z A$ to produce the derivative of the delta function.

Now, we argue that both $\m$ and $\n$ have to be $\zb$, and $\r$, $\s$ have to be along the $y^i$ directions, in all three examples studied in Section~\re{secex}.  In $f(R)$ gravity, this is trivial because no $R_{zizj}$ terms appear in the Lagrangian, and our proof is already complete with the first term \er{b1} of the brane action, which is Wald's formula.  In general four-derivative gravity \er{L4der}, the full contraction of two Riemann tensors requires that if one is $R_{zizj}$, the other must be of the form $R_{\zb k\zb l}$.  And finally in Lovelock gravity, this results from the antisymmetry of the generalized delta symbol in its Lagrangian \er{lll}.

Therefore, let us proceed in these three examples, where \er{eom2} becomes
\ba
E_{(2)}^{kl} &= -16 \na_\zb \na_\zb \( \fr{\pa^2 L}{\pa R_{zizj} \pa R_{\zb k\zb l}} K_{zij} \na_z A\) + (z\leftrightarrow \zb) \\
&= -16 \fr{\pa^2 L}{\pa R_{zizj} \pa R_{\zb k\zb l}} K_{zij} G_{z\zb} \na_\zb \tdd + (z\leftrightarrow \zb) +\cdots \,,\la{eom3}
\ea
where the dots denote terms that do not involve derivatives of $\tdd$.  It is natural to compensate this by introducing a second term of the brane action
\be\la{b2}
S_B^{(2)} = 16 \int d^D x \sqrt G \fr{\pa^2 L}{\pa R_{zizj} \pa R_{\zb k\zb l}} K_{zij} K_{\zb kl} G_{z\zb} \tdd \,.
\ee
This is the correct answer for general four-derivative gravity, because the second derivative of $L$ in \er{b2} no longer has any Riemann tensor, and the only terms involving derivatives of $\tdd$ in the equation of motion from \er{b2} is minus \er{eom3}.  This action \er{b2} is exactly $\e$ times the second term \er{ee2} in our entropy formula, once we set $q_\a=0$ for four-derivative gravity.

In general, however, \er{b2} does not work.  This is because the second derivative of $L$ may contain Riemann tensors, which produce additional terms involving derivatives acting on $\tdd$ in the equation of motion.  In Lovelock gravity, the correct prescription is to replace \er{b2} with
\be\la{b3}
S_B^{(2)} = 16 \int d^D x \sqrt G \sum_\a \( \fr{\pa^2 L}{\pa R_{zizj} \pa R_{\zb k\zb l}} \)_\a \fr{K_{zij} K_{\zb kl}}{q_\a+1} G_{z\zb} \tdd \,,
\ee
as we have already shown in Section~\re{seclb}.  We will not repeat the derivation here, except to comment on how it works from the perspective of canceling \er{eom3}.  In Lovelock gravity, the second derivative of $L$ in \er{b3} only contain Riemann tensors of the form $R_{i'k'j'l'}$.  We may determine how they contribute to the equation of motion by decomposing them according to \er{gc}.  In a full expansion, each term consists of $(q_\a+1)$ pairs of $K$ on equal footing just like discussed in Section~\re{seceell}.  Each pair of $K$ contributes the same term involving derivatives of $\tdd$ to the equation of motion, so we need to include a compensating factor $1/(q_\a+1)$ in \er{b3}.

Finally, we need to consider terms that are proportional to $\tdd$ in the equation of motion.  Note that we have fully matched the terms in \er{eom1} with the equation of motion from \er{b1}.  On the other hand, the relevant terms in \er{eom2} or \er{eom3} necessarily come with at least one power of $\r^{2\e}$ (so we can ignore them according to our matching procedure).  We can see this by simply comparing \er{eom3} with \er{eom1}, noting that the former has one fewer power of $G_{z\zb}$ (while both have two powers of $G^{z\zb}$ from the second derivative of $L$).

A similar argument holds for terms where the two derivatives $\na^\m \na_\n$ in \er{eom} act on powers of $e^{-2A}$.  In this case one can determine which components of the Riemann tensor contribute powers of $e^{-2A}$ under covariant derivatives; the answer is given by a counting argument similar to the one given below \er{rexp}.  We do not need the details here, except that the components of the Riemann tensor contribute either zero or one power of $e^{-2A}$.  Then even if a delta function is generated by $\na^\m \na_\n$ in \er{eom}, it necessarily comes with at least one power of $e^{-2A}=\r^{2\e}$, and we can ignore these terms in our matching procedure.

\se{Conclusions and discussions}\la{seccon}

In this paper, we have proposed a formula \er{ee} that, when evaluated on a well-defined surface $C_1$, gives the holographic entanglement entropy in a general theory dual to higher derivative gravity where the Lagrangian is a contraction of Riemann tensors.  Furthermore, we have shown that the surface $C_1$ is determined by minimizing the same entropy formula at least in three examples including Lovelock and general four-derivative gravity.

An immediate open question is whether minimizing our formula always gives $C_1$.  This is a very interesting question which we leave for future work.

We note that while it would be nice if minimization always works, from the point of view of calculating the holographic entanglement entropy it is not absolutely necessary to have such a minimization prescription.  The surface $C_1$ can be found by solving the bulk equations of motion with conical boundary conditions.  As long as its location can be determined by this or some other more efficient method, we can use our formula \er{ee} to calculate the entanglement entropy.  One plausible method is to always look for potentially divergent components of the equations of motion, analogous to \er{rzz} for Einstein gravity and \er{ller} for Lovelock gravity.  The resulting constraints may be sufficient to fix the location of $C_1$, even if they do not arise from minimizing some functional.  In fact, explicit computations using the Ryu-Takayanagi prescription almost always involve first converting the minimization prescription into the equations of motion for the embedding coordinates.  Hence we lose nothing in these calculations if we start with the embedding equations instead of the minimization prescription.

We also note that there is a potential difficulty even with a minimization prescription, in general cases of higher derivative gravity.  This is because the embedding equations generally involve more than two derivatives.  This difficulty may be related to the difficulty in proving the minimization prescription in general\footnote{Both of these two difficulties might be related to the fact that adding generic higher derivative terms with large coefficients to the Lagrangian often gives ghost-like excitations.}.  Lovelock gravity is particularly simple from this point of view because its equations do not have more than two derivatives.

Entanglement entropy satisfies a nontrivial inequality known as the strong subadditivity, which says $S(A)+S(B) \ge S(A\cup B) + S(A\cap B)$ for two regions $A$ and $B$.  For theories dual to Einstein gravity, we may show that the Ryu-Takayanagi prescription leads to strong subadditivity \cite{Headrick:2007km} by reconnecting the minimal surfaces for $A$ and $B$ so that they become homologous to $A\cup B$ and $A\cap B$ respectively without changing their total area.  For general higher derivative gravity, as we mentioned before a minimization prescription may not exist.  Even when it exists, reconnecting the minimal surfaces could introduce large extrinsic (and intrinsic) curvatures\footnote{We thank Aron Wall for bringing this issue to our attention.}, potentially invalidating the argument for strong subadditivity.  An exception is when the entropy formula involves only the projected curvature tensors, i.e.\ only the first term (Wald's formula) but not the second term in \er{ee}.  This is true for e.g. $f(R)$ gravity, in which case we can prove strong subadditivity using the same argument as in \cite{Headrick:2007km}.

On the other hand, we should only expect strong subadditivity when the bulk theory is a consistent theory of gravity dual to a unitary field theory.  A generic theory of higher derivative gravity with couplings that are not perturbatively small does not have to be consistent and satisfy strong subadditivity.  However, we may work perturbatively with small couplings for the higher derivative terms, and regard the theory as an effective theory approximating e.g.\ string theory (which gives an infinite tower of higher derivative interactions).  In this perturbative regime, strong subadditivity follows automatically from that of the leading area term.  It seems quite interesting to use strong subadditivity to constrain the parameter space of higher derivative couplings.

Another interesting question is how our entropy formula might be related to the approach of Iyer and Wald \cite{Iyer:1994ys}.  The Noether charge method from which Wald's formula was originally derived has ambiguities unless the horizon is bifurcate \cite{Jacobson:1993vj, Iyer:1994ys}.  For dynamical horizons, Iyer and Wald proposed the prescription of applying Wald's formula in a new geometry constructed from the old one, but in which the dynamical horizon becomes bifurcate \cite{Iyer:1994ys}.  It would be interesting to connect our formula to this setting.

Even though we have focused on the holographic entanglement entropy in this paper (as this was our original motivation), the derivation of our formula applies also for the black hole entropy, at least in the holographic case.  The formula is evaluated on the horizon which does not have to be a Killing horizon.  It would be very interesting to further explore the application of our formula in the black hole context\footnote{We thank Robert Myers for reminding us about this.}.

Let us also briefly mention another direction where our result could be useful, which is the intriguing idea that entanglement and spacetime are intimately related in some deep way in quantum gravity \cite{VanRaamsdonk:2009ar, VanRaamsdonk:2010pw, Swingle:2012wq, Bianchi:2012ev, Myers:2013lva, Balasubramanian:2013rqa, Maldacena:2013xja}.

There are several possible generalizations of our discussions.  First, we can add matter fields to the bulk theory \er{act}.  If we add only scalar fields arbitrarily coupled to the Riemann tensor, we expect that the same entropy formula \er{ee} applies with no modification.  Second, we can consider bulk Lagrangians that involve derivatives of the Riemann tensor.  We suspect that a straightforward generalization of the derivation in Section~\re{secee} should work in this case.  And finally, we may consider interesting bulk theories without diffeomorphism invariance, such as topologically massive gravity.  We leave these for future work.

\section*{Acknowledgments}
I am very grateful to Shamik Banerjee, Daniel Harlow, Sean Hartnoll, Nabil Iqbal, Juan Maldacena, Yi Pang, Stephen Shenker, Eva Silverstein, Aron Wall for discussions.  I would also like to thank especially Shamit Kachru for encouragement during early stages of this work and Robert Myers for very valuable comments.  This work was supported by the National Science Foundation under grant PHY-0756174.

\appendix

\se{Notations and conventions}\la{appnt}

In this appendix, we summarize the notations and conventions used in this paper for quick reference.

We always work in the Euclidean signature.  For example, Wald's formula \er{wald} differs from its usual Lorentzian form by a minus sign.

We use $M_n$ to denote the $n$-fold cover of the Euclidean spacetime on which the field theory is defined.  $B_n$ denotes the bulk solution dual to $M_n$, and $\hat B_n=B_n/Z_n$ is the orbifold or its analytic continuation.  The conical defect in $\hat B_n$ is called $C_n$.  We refer to the $n\to1$ limit of $C_n$ as $C_1$.

We use Greek letter $\m$, $\n$, $\r$, $\s$, $\cdots$ as indices of the $D$-dimensional bulk geometry.  We use Latin letters $a$, $b$, $\cdots$ as indices of the 2-dimensional space orthogonal to the conical defect $C_n$, and $i$, $j$, $\cdots$ as indices of the $d=D-2$ dimensional space along $C_n$.

We try to consistently use capital letters to denote higher-dimensional quantities and lower-case letters to denote lower-dimensional (intrinsic) quantities.  In particular, $G_{\m\n}$ and $R_{\m\r\n\s}$ are the $D$-dimensional metric and Riemann tensor, whereas $g_{ij}$, $\T \g{i}{jk}$, and $r_{ijkl}$ are the induced metric, intrinsic Christoffel symbol, and intrinsic Riemann tensor in $d$ dimensions.  For the 2 orthogonal directions, it is useful to strip off the singular warp factor and define $\hat g_{ab} \eq e^{-2A} G_{ab}$.

In the 2 directions orthogonal to $C_n$, we frequently switch between the cylindrical coordinates $(\r,\t)$ and the complex coordinates $(z,\zb)$, depending on which are more convenient.  Their relation is obviously $z=\r e^{i\t}$.  Note that we use $\r$ for three purposes -- the density matrix, a bulk index, and the radial coordinate -- but they are sufficiently different and have not been a cause of confusion in our experience.

We define the derivative with respect to the Riemann tensor in the standard way, so that it has the Riemann symmetry and satisfies the identity $\d L = \fr{\pa L}{\pa R_{\m\r\n\s}} \d R_{\m\r\n\s}$.

We use $K_{aij}$ to denote the extrinsic curvature tensor of the codimension 2 surface $C_n$, which is sometimes also written as $K_{(a)ij}$ or $_{(a)}K_{ij}$ in the literature.  In our coordinates it may be defined as $K_{aij} \eq \fr12 \pa_a G_{ij}$.  Its trace is written as $K_a \eq K_{aij}g^{ij}$.

If needed, we can introduce two orthogonal unit vectors $n^{(a)}_\m$ along the $x^a$ directions, and $d$ orthogonal unit vectors $m^{(i)}_\m$ along the $y^i$ directions.  In terms of these we define $n_{\m\n} = n^{(a)}_\m n^{(b)}_\n G_{ab}$ which projects to $G_{ab}$ in the $x^a$ directions.  Similarly we define $\ve_{\m\n} = n^{(a)}_\m n^{(b)}_\n \ve_{ab}$ where $\ve_{ab}$ is the usual Levi-Civita tensor, and $K_{\l\m\n} = n^{(a)}_\l m^{(i)}_\m m^{(j)}_\n K_{aij}$.

In the discussions on the cosmic brane action, we frequently use the shorthand notation $\tdd \eq 2\pi\e \d^{(2)}(x^1, x^2) / \sqrt{g^{(2)}}$ where $g^{(2)}$ is the determinant of the induced metric in the $x^a$ directions.

\se{Squashed cone metric}\la{appsc}

In this appendix, we worked out the general metric of a squashed cone which is the analytic continuation of the orbifold $\hat B_n=B_n/Z_n$.

Let us first consider the parent space $B_n$ which is defined only at integer $n$.  We may construct quasi-cylindrical coordinates $(\tdr,\tdt,y^i)$ \cite{Unruh:1989hy} within a sufficiently small neighborhood of $C_n$, such that the $Z_n$ symmetry acts by discrete rotation: $\tdt\to\tdt+2\pi/n$.  Here the range of $\tdt$ is $2\pi$.  Note that in the totally regular parent space, $C_n$ is simply the set of fixed points under $Z_n$.

By judiciously choosing the coordinates, we can write the most general metric of $B_n$ as
\bm\la{metp}
ds^2 = d\tdr^2 + \tdr^2 \[1 + \tdr^2 \mo\(1,\tdr^2,\tdr^n e^{\pm in\tdt}\)\] d\tdt^2 \\ + \[g_{ij} + \mo\(\tdr^2,\tdr^n e^{\pm in\tdt}\)\] dy^i dy^j + \tdr^2 \mo\(1,\tdr^2,\tdr^n e^{\pm in\tdt}\) d\tdt dy^i \,.
\em
Here $\mo\(1,\tdr^2,\tdr^n e^{\pm in\tdt}\)$ denotes a Taylor expansion in $\tdr^2$ and $\tdr^n e^{\pm in\tdt}$, allowing for a constant term, and $\mo\(\tdr^2,\tdr^n e^{\pm in\tdt}\)$ is the same except that a constant term is not allowed.  Note that by a Taylor expansion in $\tdr^2$ and $\tdr^n e^{\pm in\tdt}$, we mean the following systematic expansion:
\be\la{taylor}
\mo\(1,\tdr^2,\tdr^n e^{\pm in\tdt}\) \eq \sum_{k=-\infty}^\infty \(\sum_{m=0}^\infty \td c_{km} \tdr^{2m}\) \tdr^{|k|n} e^{\pm ikn\tdt} \,,
\ee
where $\td c_{km}$ are the coefficients.  Explicitly, we either have powers of $\tdr^n e^{in\tdt}$ or of $\tdr^n e^{-in\tdt}$, but not of both (choosing to ``annihilate'' them into powers of $\tdr^2$).  Right now this is just a particular way of organizing the expansion, but it will become a prescription for the analytic continuation which we will perform in a moment.

The form of the metric \er{metp} is fixed by the requirement of regularity at $\tdr=0$ and the $Z_n$ symmetry.  For example, the $Z_n$ symmetry requires that all dependence on $\tdt$ has to be in the form of $e^{\pm in\tdt}$, which must be accompanied by $\tdr^n$ for regularity.  We remind our readers that we have $G_{\tdr i}=0$ because we define the $\tdr$ coordinate by constructing orthogonal geodesics from $C_n$, and we choose the $\tdt$ coordinate so that $G_{\tdr\tdt}=0$.

We may now perform the $Z_n$ orbifold and transform to the new coordinates
\be\la{orb}
\r \eq \(\fr{\tdr}{n}\)^n \,,\qu
\t \eq n\tdt \,,
\ee
in which the metric \er{metp} becomes
\bm\la{metsc}
ds^2 = \r^{-2\e} \lt\{ d\r^2 + \r^2 \[1 + \r^{2-2\e} \mo\(1,\r^{2-2\e},\r e^{\pm i\t}\)\] d\t^2 \rt\} \\ + \[g_{ij} + \mo\(\r^{2-2\e},\r e^{\pm i\t}\)\] dy^i dy^j + \r^{2-2\e} \mo\(1,\r^{2-2\e},\r e^{\pm i\t}\) d\t dy^i \,.
\em
Here $\e\eq 1-1/n$ as before, and the range of $\t$ in the orbifold $\hat B_n$ is $2\pi$.  We now analytically continue the metric \er{metsc} to arbitrary $\e$.  For example, a Taylor expansion in $\r^{2-2\e}$ and $\r e^{\pm i\t}$ means the natural continuation of \er{taylor} after using \er{orb}:
\be\la{tayorb}
\mo\(1,\r^{2-2\e},\r e^{\pm i\t}\) \eq \sum_{k=-\infty}^\infty \(\sum_{m=0}^\infty c_{km} \r^{(2-2\e)m}\) \r^{|k|} e^{\pm ik\t} \,,
\ee
where $c_{km}$ is related to $\td c_{km}$.  In particular, we note that there is no $\r^2$ term in the expansion \er{tayorb}, but there is $\r^{2-2\e}$.  We emphasize that although normally a Taylor expansion in $\r^{2-2\e}$ and $\r e^{\pm i\t}$ might be allowed to have $\r^2$, here we define the allowed terms in \er{tayorb} by analytic continuation, and $\r^{2-2\e}$ is allowed but $\r^2$ is not.  We do not yet have a way to prove that this is the unique prescription for analytic continuation (consistent with other constraints in our problem such as the $Z_n$ symmetry), but it seems to be the simplest and most natural prescription.  Note that if the correct prescription contains terms involving additional powers of $\r^{2\e}$, this may affect the definition of $q_\a$ as presented below \er{rtd}, but as long as we define it as the number of extra $\r^{2\e}$ factors in each term in the expansion of $\fr{\pa^2 L}{\pa R_{zizj} \pa R_{\zb k\zb l}}$ (not counting those in the two powers of $G^{z\zb}$), our general formula \er{ee} holds.  We wish to revisit this issue in future work.

We may now work in complex coordinates $z\eq \r e^{i\t}$ and expand the metric as
\bm\la{met4}
ds^2 = \r^{-2\e} \[dz d\zb + \r^{-2\e} T (\zb dz-z d\zb)^2 \] + \(g_{ij} + 2K_{aij} x^a + Q_{abij} x^a x^b\) dy^i dy^j \\
+ 2i \r^{-2\e} \(U_i + V_{ai} x^a\) \(\zb dz-z d\zb\) dy^i + \cdots \,.
\em
According to \er{metsc} we find that $Q_{z\zb ij}$ secretly contains $\r^{-2\e}$, while all other coefficient tensors in the above expansion are independent of $\r$.  This was used in the counting argument below \er{rexp} in Section~\re{secan}.  Note that the squashed cone metric \er{met4} may be regulated by replacing $\r^{-2\e}$ with $e^{2A}$, yielding exactly \er{met3}.

We may also consider a scalar field $\p$.  In the parent space $B_n$ it must take the form of a Taylor expansion in $\tdr^2$ and $\tdr^n e^{\pm in\tdt}$ because of the $Z_n$ symmetry and regularity.  Therefore in the new coordinates of $\hat B_n$ it must be a Taylor expansion in $z$, $\zb$, and $\r^{2-2\e}$ in the sense of \er{tayorb}.

\bibliographystyle{JHEP}
\bibliography{references}

\end{document}